\documentclass[twocolumn,preprintnumbers,amsmath,amsfonts,amssymb,notitlepage,showpacs,aps,prb,longbibliography]{revtex4-2}

\usepackage{color}

\usepackage{dsfont}
\usepackage{amsmath}

\usepackage{graphicx}

\usepackage{hyperref}

\def \be{\begin{equation}}
\def \ee{\end{equation}}
\def \ba{\begin{array}}
\def \ea{\end{array}}
\def \bea{\begin{eqnarray}}
\def \eea{\end{eqnarray}}

\renewcommand{\textcolor}[2]{#2}

\date{\today}
\begin{document}
\title{
Meson dynamics from locally exciting a particle-conserving $Z_2$ lattice gauge theory}

\author{Vaibhav Sharma}
\email{vaibhavsharma@rice.edu}
\author{Kaden R. A. Hazzard}
\email{kaden@rice.edu}
\affiliation{Department of Physics and Astronomy, Rice University, Houston, TX 77005}
\affiliation{Smalley-Curl Institute, Rice University, Houston, TX 77005}


\begin{abstract}

Quantum simulation of lattice gauge theories is an important avenue to gain insights into both particle physics phenomena and constrained  quantum many-body dynamics. There is a growing interest in probing analogs of high energy collision phenomena in lattice gauge theories that can be implemented on current quantum simulators. Motivated by this, we characterize the confined mesons that originate from a local high energy excitation in a particle-conserving 1D $Z_2$ lattice gauge theory. We focus on a simple, experimentally accessible setting that does not require preparation of colliding wavepackets and isolates the effects of gauge field confinement strength and initial state energy on the nature of propagating excitations. We find that the dynamics \textcolor{blue}{are} characterized by the propagation of a superposition of differently sized mesons. The linear confinement leads to meson size oscillations in time. The average meson size and oscillation frequency are controlled by the strength of the gauge field confinement. At a constant confinement field, the average meson length is controlled by the initial excitation's energy. Higher energies produce longer mesons and their effective mass depends strongly on their size: longer mesons propagate more slowly out of the central excitation. Mesons of different sizes get spatially filtered with time due to different speeds. We show that this phenomenology is a consequence of linear confinement and remains valid in both the strong and weak confinement limit. We present simple explanations of these phenomena supported by exact numerics.

\end{abstract}

\maketitle

\section{Introduction}\label{intro}

Gauge theories model vast phenomena, ranging from the interactions of fundamental particles to the collective behaviour of spins in condensed matter systems~\cite{kogut}. Over the past decades, we have learnt a great deal about gauge theories like QED and QCD through large-scale particle collider experiments where high-energy particles are smashed together. These high energies can give rise to new particles and interesting non-equilibrium phenomena such as a quark-gluon plasma and short-lived bound states~\cite{qgplasma}. However, it is computationally challenging to access the real-time non-equilibrium phenomena in these collisions experiments using quantum Monte Carlo lattice QCD techniques on classical computers. 

Quantum computers and quantum simulators can bridge this gap. By implementing these gauge theories on a discrete lattice, \textcolor{blue}{error corrected quantum computers would be able to simulate real-time quantum dynamics of high-energy phenomena to high levels of precision. Current quantum processors can already implement dynamical simulation experiments of some simple models such as $Z_2$ and $U(1)$ lattice gauge theories, albeit with significant noise effects~\cite{exp1,exp2,exp3,exp4,exp5,exp6,exp7,exp8,exp9,stringbreaking3,Gorshkov6,Dalmonte4}.}  

Although these simple models are far from the more complicated SU(3) QCD in three dimensions, they display some of the features we expect in QCD such as linear confinement of particles, pair production, and string breaking~\cite{stringbreaking,stringbreaking2,stringbreaking3,stringbreaking4}. Thus studying such simple models already can give us insights into the dynamical phenomena we expect in gauge theories. It also gives us a roadmap where we can leverage current quantum simulator technology to understand simple lattice gauge theory models and gradually build-up towards a full simulation of gauge theories like QCD. In addition to the connection to high energy physics, these gauge theory models also emerge as descriptions of low energy phases in condensed matter systems such as topologically ordered quantum spin liquids~\cite{Zoller1,Zoller2,Gorshkov1,Grusdt1,Lewenstein3,toporder}. The interplay of gauge symmetry and strong interactions also leads to a rich set of quantum many-body phenomena~\cite{sectors,manybody1,manybody2,manybody3,manybody4,manybody5,manybody6}. In the condensed matter context, it is important to note that these theories genuinely live on a lattice, as opposed to the lattice being introduced to approximate continuous space, leading to unique phenomena.

The study of dynamical phenomena in lattice gauge theories is attracting considerable attention in recent years. These gauge theories show exotic non-ergodic phenomena such as lack of thermalization due to confinement and Hilbert-space fragmentation~\cite{Halimeh1,Halimeh7,Gorshkov7,Lewenstein4,exp6,bessel}, quantum many-body scars~\cite{Halimeh6,Halimeh8,Lewenstein1,Lewenstein2} and Floquet time crystals~\cite{Dalmonte2}.  Analogs of QCD phenomena such as confinement dynamics~\cite{Calabrese1,Dalmonte4,Gorshkov6,fractonconfinement}, pair production through string breaking~\cite{Dalmonte1,Dalmonte3,stringbreaking4,Halimeh5,Gorshkov5,dynamicallocalization,stringbreaking,stringbreaking2,stringbreaking3}, false vacuum decay~\cite{Halimeh3}, and dynamical hadron formation~\cite{hadron1,hadron2} have been found in theoretical calculations and some quantum simulator experiments.

These successes have opened up an exciting new avenue of probing analogs of high energy particle collision phenomena within these simple lattice gauge theories. Some recent works have studied schemes to prepare moving wavepackets for collisions \textcolor{blue}{~\cite{collider,scattering,scattering2} and numerically analyzed scattering processes~\cite{scattering2,Gorshkov3,Gorshkov4,hadron1}.} In these studies, the dynamics involves a rich interplay of multiple kinds of bound excitations, resonant processes of particle pair production and string fluctuation dynamics. 

This rich complexity can mask the fundamental role played solely by gauge-mediated interactions. For instance, in the absence of resonant string breaking and particle production processes, it is not fully clear how the nature of the propagating bound state excitations systematically varies with the strength of the confining field and the magnitude of a localized energy injection event.  Through this work, we intend to understand such building blocks of collision phenomena in lattice gauge theories in a simpler experimentally accessible setting and probe the effects of confinement strength and initial energy.

In our work, we characterize the nature of propagating excitations that emerge from a localized high energy excitation in a particle-conserving 1D $Z_2$ lattice gauge theory model. A schematic of the scenario we study is shown in Fig.~\ref{fig:latticediag}(a) in which a high energy excitation disperses into outward propagating bound mesons. In a collider experiment, such a high energy region occurs at the location of a collision. However in a controlled quantum simulation, a high energy excitation region can also be created without a collision.
We consider starting with product states where a small local region is excited to a tunable higher energy using local gauge spin rotations as opposed to collisions of moving wavepackets. Our setting provides an alternate scenario of creating non-equilibrium excitations from an initial energy injection.

In particular, we analyze the excitations produced by initially localizing two particles in the center of a 1D chain in its vacuum state. Our model conserves particle number and shows confinement of two particles into mesons of different possible lengths~\cite{confinement}. We use a mix of simple analytical calculations and exact diagonalization numerics to characterize the structure of mesons and their propagation speed as they emerge from this excited region. 

We find that the long time state is characterized by propagation of a superposition of differently sized mesons, with the average size inversely dependent on the confinement field. The meson size oscillates in time and the oscillations get larger as the confinement field reduces. These larger mesons slowly propagate out, damping out the oscillations near the center. Some of this phenomenology is similar to what has been seen in wavepacket scattering studies in lattice gauge theories where a particle-anti particle pair after a collision continues to periodically move apart and come back for another collision~\cite{Gorshkov4,collider}. The scaling of the average meson size and the oscillation frequency are qualitatively described by analytical results obtained from an associated Wannier-Stark problem, which becomes exact as we approach infinitely large confinement field strengths.

We also consider the effect of tuning the initial state energy at a fixed confinement field. The average length of the meson that propagates monotonically increases as a function of initial state energy, instead of the spatial size of the initial excitation. As the meson size increases, its propagation speed is slower, which slows down the spread of particle density away from the excitation region. \textcolor{blue}{Since the long time wavefunction of even a single meson is a superposition of differently sized mesons, the correlation between meson size and speed causes a size dependent spatial separation of the mesons participating in the superposition.} This is a consequence of linear confinement and we find that this feature holds true even for cases where the confining field is weaker than the particle hopping term. We show that for weaker confining fields, mesons longer than a certain size experience a slow down. 

\section{Model}

We consider hard-core bosons occupying the sites of a one-dimensional lattice, coupled to spin-1/2 degrees of freedom on the links of the lattice. The 1D $Z_2$ lattice gauge theory Hamiltonian we study is

\begin{equation}\label{hamiltonian}
    {H} = -J \sum_{i} (b_{i+1}^{\dag}\sigma^{x}_{i,i+1} b_{i} + h.c.) + h\sum_{i} \sigma^{z}_{i,i+1}.
\end{equation}

\begin{figure}
    \includegraphics[width=\columnwidth]{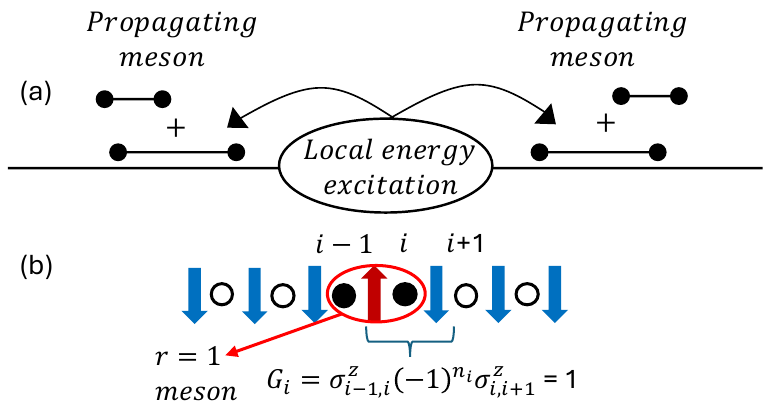}
    \caption{(a) \textcolor{blue}{Meson dynamics: A meson propagates out} of the localized energy excitation in both directions in a 1D chain governed by a $Z_2$ lattice gauge theory. The state is described by a superposition of different meson sizes. The confinement field and the initial excitation energy determine the meson size distribution. The propagation speed varies with the meson size. (b) A basis state of the 1D $Z_2$ lattice gauge theory model. Open and filled circles represent absence and presence of a boson on sites labeled by index $i$. The red and blue arrows are link gauge spins representing $\sigma^z = \pm 1$ respectively. The quantity, $G_i = \sigma^z_{i-1,i}(-1)^{n_i}\sigma^z_{i,i+1} = 1$ is conserved on each site. The red ellipse denotes a meson where the two bosons are next to each other, called an $r=1$ meson.}
    \label{fig:latticediag}
\end{figure}

The operators $b^{\phantom \dagger},b^{\dagger}$ annihilate and create hardcore bosons on site $i$ and $\sigma^x,\sigma^z$ are spin-1/2 Pauli operators for the link degrees of freedom. The bosons are considered as the matter particles while the spin-1/2 particles on the links are considered to be the gauge fields. A basis state of the model is shown in Fig.~\ref{fig:latticediag}(b). 

Bosons hop between nearest-neighbor sites by flipping a link spin. This is the $Z_2$ analog of minimal coupling that couples the matter's kinetic energy to the gauge field vector potential. In this analogy, $\sigma^x_{i,i+1} = e^{iA_{i,i+1}}$ where $A_{i,i+1} = \pi/2 (\sigma^x_{i,i+1}-1)$. The electric field operator conjugate to the vector potential operator $A_{i,i+1}$ is given by $\sigma^z_{i,i+1}$. Thus the field term proportional to $h \sigma^z_{i,i+1}$ is the energy cost of the electric field due to excitations of the gauge field. 

The model in Eq.~\ref{hamiltonian} describes  a $Z_2$ gauge theory since the Hamiltonian is invariant under a local $Z_2$ gauge transformation generated by the operator, $G_i = \sigma^{z}_{i-1,i}(-1)^{n_i}\sigma^{z}_{i,i+1}$. Here $n_i = b_{i}^{\dag} b_i$ denotes the number of bosons on site $i$. Under $G_i$, the operators in the Hamiltonian undergo $Z_2$ transformations, $G_i b_i G_i^{\dagger} = e^{i \pi} b_i$, $G_i b_i^{\dagger} G_i^{\dagger} = e^{-i \pi} b_i^{\dagger}$ and $G_i \sigma^x_{i,i+1} G_i^{\dagger} = e^{i \pi} \sigma^x_{i,i+1} $.

There is a local conservation law at each site since $[G_i,H] = 0$. The Hamiltonian eigenstates can thus be chosen to also be the eigenstates of the operator $G_i$ at each site. The eigenvalues of $G_i$ are $\pm 1$. Different choices of the $G_i$ eigenvalues correspond to independent and physically distinct sectors of the Hamiltonian's Hilbert space~\cite{sectors}. Throughout this work, we only consider states $|\psi\rangle$ where $G_i|\psi\rangle = +1|\psi\rangle$ for all sites $i$. This constraint implies that $\sigma^{z}_{i-1,i}(-1)^{n_i}\sigma^{z}_{i,i+1}|\psi\rangle = |\psi\rangle$. We can rewrite it as, $\sigma^{z}_{i-1,i}\sigma^{z}_{i,i+1}|\psi\rangle = e^{i\pi/2(\sigma^z_{i-1,i}-\sigma^z_{i,i+1})} =(-1)^{n_i}|\psi\rangle$. This can be interpreted as a Gauss law where the local matter density determines the electric field configuration around it.

\section{The two-particle sector}\label{sec:twoparticle}

The Hamiltonian in Eq.~\ref{hamiltonian} conserves total boson number. In this work, we consider dynamics when there are only two bosons present in the system. We consider configurations where the two bosons are added onto the vacuum, the particle-free ground state. The vacuum state has all the gauge spins aligned along the negative $z$-axis ($\sigma^z_{i,i+1} = -1$) to minimize the electric field energy (we assume $h>0$). An example of such a two-particle configuration is shown in Fig.~\ref{fig:latticediag}(b). Due to the Gauss law constraint on each site, the link spin configuration can be completely determined by the position of the bosons. We thus get, $\sigma^z_{i,i+1}=\prod_{j\leq i} (-1)^{n_j}$. 
The states in the two-boson sector can be characterized by the position indices of the two bosons, $i_1$ and $i_2$. The link spins between sites $i_1$ and $i_2$ are the only ones aligned in the positive $z$-direction leading to an electric field energy cost of $+2h$ for each such link. This is what leads to confinement of particles into mesons in this model.

Using the two bosons' positions ($i_1,i_2$) to label the basis states, we can write down the Hamiltonian in Eq.~\ref{hamiltonian} in the two-boson sector as
\begin{multline}\label{hamiltonian1}
    H = -J\sum_{\textcolor{blue}{i_1 \neq i_2}} (|i_1+1,i_2\rangle\langle i_1,i_2| + |i_1,i_2 + 1\rangle\langle i_1,i_2| + \text{h.c.} ) \\
    + 2h \sum_{\textcolor{blue}{i_1 \neq i_2}} |i_1-i_2||i_1,i_2\rangle \langle i_1,i_2|.
\end{multline}

\textcolor{blue}{Due to the hard-core limit for the bosons, the position coordinates for the two bosons must be unequal}. We use the center-of-mass $(c = (i_1+i_2)/2)$ and relative $(r = i_1-i_2)$ coordinates to further simplify the Hamiltonian. The Hamiltonian in Eq.~\ref{hamiltonian1} becomes
\begin{equation}\label{hamiltonian2}
    \begin{split}
      H &= -J\sum_{c}\sum_{r=1}^{\infty} (|c+1/2\rangle\langle c| + \text{h.c.})(|r+1\rangle\langle r| + \text{h.c.}) \\
    &+ 2h \sum_{r=1}^{\infty} r|r\rangle \langle r|  
    \end{split}
\end{equation}

Here the center of mass coordinate $c$ can take half-integer and integer values while the relative coordinate $r$ only takes integer values \textcolor{blue}{where $r > 1$ due to the hardcore boson limit}. The relative coordinate has a linear potential term in Eq.~\ref{hamiltonian2}. In the ground state, at any finite $h/J$, two particles are confined by this potential into mesons~\cite{confinement}. The center-of-mass coordinate is free and thus the mesons can delocalize. In this work, we study the two-particle sector dynamics in terms of the center-of-mass and relative coordinates. The different relative coordinate values correspond to different types of possible meson excitations. The smallest possible meson is the $r=1$ meson where the two bosons are right next to each other. In later sections, we calculate how the size of the propagating mesons depends on field term $h/J$ and the energy of the initial local excitation. We will also determine how meson size affects its propagation speed. 

\subsection{Diagonalizing the Hamiltonian}\label{sec:diagonalization}

We can diagonalize the Hamiltonian in Eq.~\ref{hamiltonian2} by considering the eigenstates to be of the form, $|\psi_{n,k}\rangle = \Phi_k \otimes \Gamma_n = (\sum_c \phi^k_c |c\rangle)\otimes(\sum_r \gamma^n_r |r\rangle) $ that satisfy the equation, $H |\psi_{n,k}\rangle = E_{n,k}|\psi_{n,k}\rangle$. Note that the center-of-mass states only have a hopping term in the Hamiltonian. Thus plane wave states of the form, $\phi^k_c = e^{ikc}/\sqrt{2L-1}$ are eigenstates for the center-of-mass degree of freedom. Here $L$ is the total number of sites in the chain. Due to the half integer steps of the coordinate $c$, the momentum $k$ ranges from $-2\pi$ to $2\pi$ in steps of $2\pi/(2L-1)$. We can find the function $\gamma^n_r$ by substituting $\phi^k_c$ into the eigenvalue equation of the Hamiltonian in Eq.~\ref{hamiltonian2} and projecting it onto $|c\rangle \otimes |r\rangle$. We then get 
\begin{equation}\label{wannierstark}
    E_{n,k}\gamma^n_r = -2 J\cos(k/2)(\gamma^n_{r+1} + \gamma^n_{r-1})
    +  2h r \gamma^n_r.
\end{equation}
    
Note that the relative coordinate is restricted to $r \geq 1$.  Without such a restriction, this describes a Wannier-Stark ladder and Eq.~\ref{wannierstark} admits Bessel function solutions where $\gamma^n_r = J_{r-n}(2J \cos(k/2)/h)$~\cite{Wannierstark}. The energy eigenvalue in that case is given by, $E_{n,k} = 2hn$ where $n$ is integer quantized. The energy is independent of the center-of-mass momentum leading to a large degeneracy. However in our case, $r \leq 0$ is not allowed and thus we must enforce the boundary condition that $\gamma^n_{r=0} =0$. This is akin to adding an infinite potential wall at $r=0$. \textcolor{blue}{For a semi-infinite chain with a hard wall at $r=0$, the eigenstate solutions are again Bessel functions where $n$ is no longer integer quantized~\cite{bessel,scattering2}. The value of $n$ is found from the condition that at $r=0$, we must have $J_{-n}(2J \cos(k/2)/h) = 0$. For finite values of $h/J$, we would find a non-integer momentum dependent value of $n$ such that $2J \cos(k/2)/h$ is a zero of the Bessel function.  The energy bands would thus no longer be flat with respect to momentum $k$. When $h/J \to \infty$, $n$ tends towards integer values independent of $k$ and we recover the flat Wannier-Stark bands. If the chain is finite with hard wall boundaries at both ends, the exact eigenstates are given by Lommel polynomials and the eigenenergies are obtained from the zeros of the Lommel polynomial~\cite{lommel}.}

Even with these boundary conditions, the eigenstates and eigenenergies are analytically rather simple in the large-$h/J$ and small-$h/J$ limit. These limits essentially capture a complete picture of the physics. We now derive the solution in these two simple limits. 

\subsubsection{Large-$h/J$ limit}
When $h/J \gg 1$, the Bessel function solutions with integer quantized eigenenergies approach the exact eigenstates. At large $h/J$, the asymptotic form of the solutions is, $J_{r-n} \sim (2J\cos (k/2)/h)^{|r-n|} $. \textcolor{blue}{Since $h/J \gg 1$, the amplitude shows a decay with an increasing value of the exponent, $|r-n|$.}
The amplitude at $r=0$ is $\sim (2J\cos (k/2)/h)^{\textcolor{blue}{|n|}}$. As $h/J \to \infty$,  when $n \geq 1$, this amplitude goes to zero, satisfying the boundary condition in the asymptotic limit. The energy eigenvalues are quantized in this limit and given by, $E_{n,k} = 2hn$. The $n^{\text{th}}$ eigenstate is sharply peaked around $r=n$ with power law tails ($1/h^{|r-n|}$). Seen another way, the zeroth order $h/J \to \infty$ solution has the $n^{\text{th}}$ eigenstate as a delta function on $r=n$ with the eigenenergy $2hn$ where $n \geq 1$. The hopping term proportional to $J$ in Eq.~\ref{wannierstark} provides first order perturbation to the wavefunction and gives it a tail such that the amplitude on level $r \neq n$ is $(2J\cos (k/2)/h)^{|r-n|}$.  

\subsubsection{Small-$h/J$ limit}

When $|h/J| \ll 1$, we can take a continuum limit of the eigenvalue problem in Eq.~\ref{wannierstark} to get 

\begin{equation}\label{airy}
    E_{n,k}\gamma^n_r = -2J\cos (k/2) \frac{d^2\gamma^n_r}{dr^2} + 2hr\gamma^n_r
\end{equation}

We define $r' = l_0 r$ where $l_0 = (J\cos (k/2)/h)^{1/3}$ and rewrite Eq.~\ref{airy} as

\begin{equation}\label{airy1}
    -\frac{d^2\gamma^n_r}{dr'^2} + (r'-E'_{n,k})\gamma^n_r=0
\end{equation}
with $E'_{n,k} = E_{n,k}/2(Jh^2\cos (k/2))^{1/3}$. Eq.~\ref{airy1} is solved by Airy functions of the first kind. Thus $\gamma^n_r = \mathrm{Ai}(r'-E'_{n,k})$. The Airy functions decay exponentially to zero as $r' \to \infty$. The boundary condition at $r=0$ implies that, $\gamma^n_0 = \mathrm{Ai}(-E'_{n,k}) = 0$. The scaled eigenenergy $E'_{n,k}$ is determined by the location of the $n^{\text{th}}$ zero of the Airy function. The eigenenergy becomes, $E_{n,k} = -2z_n (Jh^2\cos (k/2))^{1/3}$, where $z_n$ is the location of the $n^{\text{th}}$ zero; these zeros are $z_1 = -2.33811, z_2 = -4.08795, z_3 = -5.52056$ and so on. Note that in this small $h/J$ limit, the eigenenergies do depend on the center-of-mass momentum and have a non-linear dependence on the confining field $h$. 

\textcolor{blue}{At intermediate values of the confining field $h/J$, one needs to find the exact momentum dependent eigenenergies using the zeros of the non-integer order Bessel functions~\cite{bessel,scattering2}.} 

\section{Effect of confinement field on meson dynamics}\label{sec:fieldeffect}

In this section, we characterize the meson dynamics from simple initial states as a function of confinement field strength. We first consider the initial excitation where an $r=1$ meson is localized at the center of a 1D chain of length $L$. This corresponds to the initial configuration shown in Fig.~\ref{fig:latticediag}(b). In the $|r,c\rangle$ (relative and center-of-mass coordinate) basis, the wavefunction at time $t=0$ is given by $|\psi(0)\rangle = |1,(L+1)/2\rangle$.  Under the Hamiltonian in Eq.~\ref{hamiltonian2}, the system wavefunction evolves in time and can be written as, $|\psi(t)\rangle = \sum_{c,r} g_{c,r}(t)|c,r\rangle$. In our calculations, we fix system size to $L=100,120$ and use exact diagonalization to calculate the time-evolved wavefunctions. Appendix~\ref{app:methods} discusses details of this calculation.

\begin{figure*}
    \includegraphics[width=15cm]{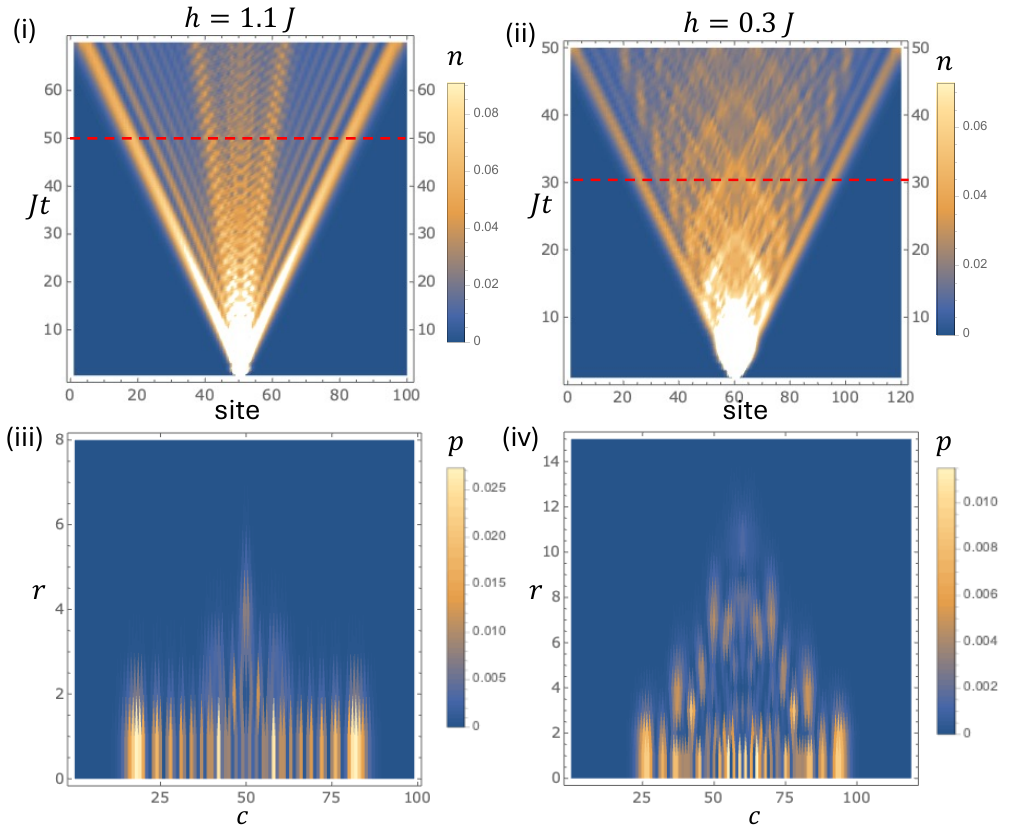}
    \caption{(i),(ii): Particle density, $n$ as a function of site index (x-axis) and time, $Jt$ (y-axis) for (i) $h/J = 1.1$ and (ii) $h/J = 0.3$. With time, the meson delocalizes in light-cones like structures. The outermost cones represent faster moving excitations. The slower moving inner cones show substructures with diamond shaped particle tracks denoting temporal oscillation of meson size. (iii),(iv): Snapshot of occupation probability, $p$ as a function of meson size ($r$) and its center-of-mass position ($c$) at (iii) $Jt = 50$ for $h=1.1J$ and (iv) $Jt = 30$ for $h=0.3J$. The red dashed lines in (i),(ii) denotes these two chosen time steps. In both cases, smaller sized mesons delocalize the farthest, suggesting larger mesons propagate slower. The different speeds lead to a spatial filtering of meson sizes as such that it is more probable to detect a longer meson closer to the center during time evolution. \textcolor{blue}{Note that we have removed zero density pixels to correct for a parity effect in (iii),(iv) because when $r$ is even, the center-of-mass $c$ only takes integer values while for odd $r$, it only takes half-integer values.}}
    \label{fig:densityplot}
\end{figure*}

We show the dynamics of propagating excitations in Fig.~\ref{fig:densityplot} at two values of the confining field, $h = 1.1J$ and $h=0.3 J$. In Fig.~\ref{fig:densityplot}(i),(ii), \textcolor{blue}{we plot the expectation value of the particle number, $n$, as a function of site index and time, $Jt$.} The particles delocalize through light-cone like structures emanating from the central excitation region at time $Jt=0$. The outermost cones in both cases represent fast moving excitations. The inner cone has a slower spread and has substructures that show diamond shaped particle tracks that also slowly spread out from the center. These are more pronounced and wider for smaller $h = 0.3J$. These tracks show similarities to the ones found after wavepacket collisions in Ref.~\cite{Gorshkov4,collider} where they were interpreted as the two particles coming together and colliding multiple times. We will show in later sections that we can interpret them as meson size oscillations which gradually spread out from the center as these mesons delocalize. \textcolor{blue}{Similar oscillations were also observed in studies involving Hamiltonian quench dynamics in Ising models~\cite{mesonoscillation,Gorshkov6,Calabrese1}.}
In Fig.~\ref{fig:densityplot}(iii),(iv), \textcolor{blue}{we plot the occupation probability, $p$, of a meson of length $r$ occupying a center-of-mass position $c$.} We can clearly see that at a given time, 
\textcolor{blue}{the largest center-of-mass positions away from the central excitation region have the highest probability of occupation from a smaller sized meson.} This suggests that meson size is inversely related to propagation speed. \textcolor{blue}{As the long time wavefunction of a single meson is characterized by a superposition of different meson sizes, this speed differential causes spatial separation of the mesons involved in the superposition according to their size.} We will show that this is a consequence of the linear confinement of mesons and persists for both $h>J$ and $h<J$ regimes.

We use the system wavefunction at long times to characterize \textcolor{blue}{the propagating excitation.} In particular, we calculate the average meson size, $r_{\text{avg}}(t) = \sum_{r} r|\langle r|\psi(t)\rangle|^2$ as a function of time. The meson \textcolor{blue}{excitation propagates} equally in both the directions away from the center. We determine the speed at which the \textcolor{blue}{excitation propagates} by calculating the temporal slope of the center-of-mass coordinate's standard deviation, $c_s$, defined as

\begin{equation}
    c_s(t)^2 = \sum_{c} c^2 |\langle c|\psi(t)\rangle|^2 - \left(\sum_{c} c |\langle c|\psi(t)\rangle|^2 \right)^2
\end{equation}

At time $t=0$, the average center-of-mass coordinate is at $(L+1)/2$ with $c_s=0$. As the \textcolor{blue}{excitation propagates} in both directions, the center-of-mass coordinate's distribution widens around the initial state average. The width is characterized by $c_s$ and is expected to increase as a function of time. Its slope gives the speed of propagation.

\begin{figure}
    \includegraphics[width=\columnwidth]{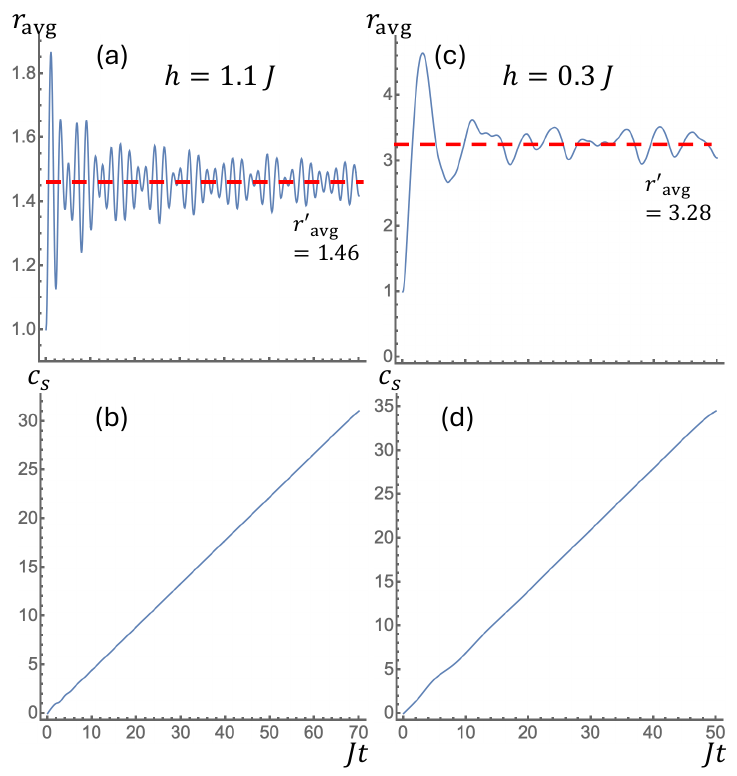}
    \caption{Dynamics after initializing an $r=1$ meson at the center of a chain of $L=100$ and $L=120$ sites for $h=1.1J$ \textcolor{blue}{and $h=0.3J$ respectively.} (a,c) Average size of propagating mesons, $r_{\text{avg}}$, vs time, $t$. The meson size oscillates in time about a value of $r'_{\text{avg}} = 1.46$ and  \textcolor{blue}{$r'_{\text{avg}} = 3.28$ for $h/J=1.1J$ and $h=0.3J$ respectively.} (b,d) Width, $c_s$, of the center-of-mass distribution vs time, $t$. Due to the propagating meson, the width of the center-of-mass distribution increases monotonically with time. \textcolor{blue}{The width grows faster for lower value of $h/J$.}}
    \label{fig:dynamics}
\end{figure}

\textcolor{blue}{In Fig.~\ref{fig:dynamics}, we show $r_{\text{avg}}$ and $c_s$ as a function of time for two values of $h/J$, $h=1.1J$ (Fig.~\ref{fig:dynamics}(a,b)) and $h/J =0.3$ Fig.~\ref{fig:dynamics}(c,d). After some transient behavior, $r_{\text{avg}}$ settles and oscillates around a steady value of $r'_{\text{avg}} = 1.46$ and  $r'_{\text{avg}} = 3.28$ for $h=1.1J$ and $h=0.3J$ respectively.} This suggests that the \textcolor{blue}{propagating excitation is a} superposition of different meson sizes. We also see that $r_{\text{avg}}$ oscillates in time, suggesting meson size oscillations. \textcolor{blue}{The oscillation amplitude is expected to asymptote to a constant value in the infinite time limit. A slight decay of this amplitude even at longer times in our plots is a consequence of our finite time simulations. The finite system size in our simulation restricts us from going to longer times as the particles start hitting the boundary. }

The width of center-of-mass distribution $c_s$ steadily increases from $c_s=0$ as the excitation propagates outwards from the center. This is expected since confined mesons can freely delocalize. \textcolor{blue}{This delocalization is faster for $h=0.3J$ compared to $h=1.1J$.} The average size of the \textcolor{blue}{propagating meson} and frequency of size oscillations are strongly dependent on the confining field, $h/J$.

\begin{figure}
    \includegraphics[width=\columnwidth]{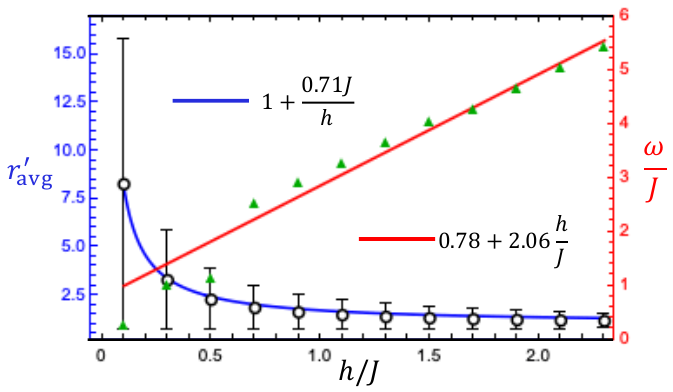}
    \caption{Long-time average size of the propagating meson, $r'_{\text{avg}}$, vs confinement field, $h/J$ shown by open markers. The solid blue line is a fit to a function, $r'_{\text{avg}} - 1 \propto 1/h$. \textcolor{blue}{The error bars show the long time average of the standard deviation of the meson size, $r'_{\text{sd}}$. It shows the range of meson size fluctuations.} 
    Frequency of oscillations, $\omega$ of average meson size, $r_{\text{avg}}$ vs field $h/J$ shown by solid green triangles. The solid red line is a fit of $\omega$ as a linear function of $h$.}
    \label{fig:sizefreq}
\end{figure}

In Fig.~\ref{fig:sizefreq}, we plot $r'_{\text{avg}}$ as a function of the confinement field $h/J$. This is the value around which $r_{\text{avg}}$ oscillates in time. We find that as $h/J$ decreases, the \textcolor{blue}{propagating meson excitation gets} longer such that $r'_{\text{avg}} -1 \propto \textcolor{blue}{J/h}$. \textcolor{blue}{The error bars show the long time average of the standard deviation of meson size, $r_{\text{sd}}'$. We first find the standard deviation of meson size, $r_{\text{sd}}(t) = \sum_r \sqrt{r^2 |\langle r|\psi(t)|^2 - r_{\text{avg}}(t)^2}$. The long time average of $r_{\text{sd}}(t)$ gives $r_{\text{sd}}'$. As $h/J$ decreases, this quantity increases as seen in the larger error bars in Fig.~\ref{fig:sizefreq}, suggesting larger fluctuations in meson size. }

We also extract the frequency of oscillations of $r_{\text{avg}}$ at long time. We take the Fourier transform of the  $r_{\text{avg}}$ vs time data at long times and use the frequency of the dominant peak as the oscillation frequency.  Fig.~\ref{fig:sizefreq} shows this frequency of oscillations, $\omega$ as a function of $h/J$. We find that $\omega \propto h$ for large $h/J$ and deviates for smaller $h/J$. \textcolor{blue}{Note that as $h/J$ reduces, the finite size effects become more important as the meson undergoes large fluctuations in size with faster delocalization towards the boundary of our system.}

These features can be qualitatively understood from Eq.`\ref{wannierstark} where the relative coordinate (meson size) moves in a Wannier-Stark ladder of a linearly tilted potential. For moderate values of $h/J$, the relative coordinate can take values away from the boundary ($r=1$) and the energy is nearly quantized, independent of the center-of-mass momentum. In this scenario, the relative coordinate undergoes breathing oscillations around its initial value. These breathing oscillations quantify the spread in meson size as a function of time. In Appendix~\ref{appendix1}, we show that the breathing oscillation function is given by, $r_s = \sqrt{2}J/h |\sin (ht)|$. The oscillation frequency of this breathing is given by, $\omega = 2h$ showing the $\omega \propto h$ dependence. The amplitude of $r_s$ sets the spread of meson sizes and controls
the inverse $h/J$ scaling of the average meson size, $r'_{\text{avg}}$. For smaller $h/J$, the meson size oscillations become larger and higher length mesons obtain finite probability of occupation, as can be seen in Fig.~\ref{fig:densityplot}(ii),(iv). 

When $h/J \gg 1$, the relative coordinate is restricted to values around $r=1$. \textcolor{blue}{We note that in this limit, the initial state is nearly an eigenstate of the hamiltonian and remains stationary for long times. However there would still be small fluctuations in the mean value of the relative coordinate as a function of time.} In Appendix~\ref{appendix2}, we show $r_{\text{avg}}= 1 + J^2/(2h^2) + J^2\sin^2(ht)/(2h^2)$. We again get frequency of oscillations, $\omega = 2h$ but the spread of average meson size around its initial value ($r=1$) scales as $(h/J)^{-2}$. For very large values of $h/J$, the curve is expected to follow an $(h/J)^{-2}$ dependence.  

When $h/J \ll 1$, the eigenfunctions are better described by Airy functions where the energy is a non-linear function of both the center-of-mass momentum and $h/J$ (see Sec.~\ref{sec:diagonalization}).
In terms of the field dependence, we expect the frequency $\omega$ to scale as $(h/J)^{2/3}$ and $r'_{\text{avg}}$ to scale as $(h/J)^{-1/3}$. Due to the special boundary condition of the relative coordinate, scalings can be more easily predicted compared to prefactors.

\textcolor{blue}{We note that the $(h/J)^{-1}$ dependence of $r'_{\text{avg}}$ (seen in Fig.~\ref{fig:sizefreq}) only captures an overall trend for moderate values of $h/J$. This scaling connects the analytically expected scalings in the $h/J \ll 1$ and $h/J \gg 1$ limits. Our numerical simulations focus on these moderate $h/J$ values as this is an interesting non-trivial regime where the simple analytical approximations presented earlier break down. In Appendix~\ref{app:scaling}, we show a log-log plot of $(r'_{\text{avg}}-1)$ vs $h/J$ where we show how the scaling changes from $(h/J)^{-1}$ to $(h/J)^{-2}$ as we move towards much larger values of $h/J$.}  
 
%



\section{Effect of the initial state's energy on meson dynamics}

\begin{figure}
    \includegraphics[width=\columnwidth]{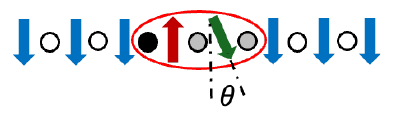}
    \caption{A central excitation in a superposition of an $r=1$ and $r=2$ meson, tuned by the angle $\theta$ of one of the gauge spins. The gray circles show a finite probability between $0$ and $1$ of having a particle controlled by $\theta$. The excitation energy is, $E(\theta) = \sin \theta + 2h \sin^2 (\theta/2) + 2h$. }
    \label{fig:r2meson}
\end{figure}

At fixed values of the confining field $h/J$, we can tune the nature of excitations that emerge by changing the initial excitation energy. This is akin to collider experiments where changing the center-of-mass collision energy determines what excitations can be produced. We consider adding some amplitude of $r=2$ meson to the $r=1$ meson initial state in order to tune the energy as shown in Fig.~\ref{fig:r2meson}. Such a configuration corresponds to tilting one of the gauge spins by an angle $\theta$ about the $y$-axis. In the $|r,c\rangle$ basis, such initial states are
\begin{equation}
    |\psi_{\theta}(0)\rangle = \cos {\frac{\theta}{2}}|1,(L+1)/2)\rangle - \sin \frac{\theta}{2} |2,L/2+1\rangle.
\end{equation}

\begin{figure}
    \includegraphics[width=\columnwidth]{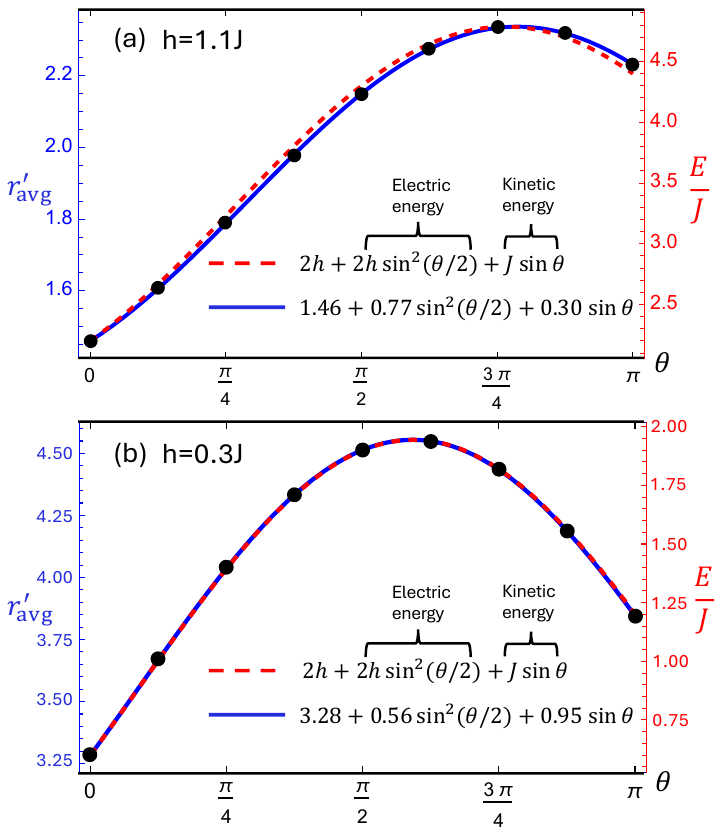}
    \caption{Initial state's energy, $E$ (red dashed line) and average meson size, $r'_{\text{avg}}$ (black dots are data and blue solid line is a fit) as a function of $\theta$ for (a) $h/J = 1.1$ \textcolor{blue}{and (b) $h/J = 0.3$.} The meson size and energy have the same functional dependence on $\theta$ in both cases. Total energy has two contributions, kinetic energy from the hopping term and the electric energy from the gauge field confinement term. Higher total energy leads to longer meson excitation that propagates.}
    \label{fig:energysizevstheta}
\end{figure}

Using the Hamiltonian in Eq.~\ref{hamiltonian2}, the energy of this initial state is, $E(\theta) = \textcolor{blue}{J} \sin \theta + 2h \sin^2 (\theta/2) + 2h$. The term $2h \sin^2 (\theta/2)$ comes from the confining electric field energy of the $r=2$ meson amplitude and the $\sin \theta$ term denotes the kinetic energy from the hopping term of the Hamiltonian. We tune $\theta$ from $0$ to $\pi$, varying the initial state from an $r=1$ meson continuously to an $r=2$ meson. \textcolor{blue}{The energy, $E(\theta)$, varies non-monotonically with $\theta$ as shown in Fig.~\ref{fig:energysizevstheta}(a),(b) for $h/J=1.1,0.3$.} This is due to the kinetic energy contribution. Starting from $|\psi_{\theta}(0)\rangle$ at a fixed confining field, we calculate the average size, $r'_{\text{avg}}$, of the \textcolor{blue}{propagating meson at long times and plot it in Fig.~\ref{fig:energysizevstheta}(a),(b) as a function of $\theta$ for $h/J=1.1$ and $h/J=0.3$ respectively.} We find that $r'_{\text{avg}}$ closely tracks the initial state's energy in both cases, having the same functional dependence on $\theta$ as energy does. If the initial energy increases, $r'_{\text{avg}}$ rises, indicating a larger meson. Note that although the initial state is spatially largest when $\theta = \pi$ (an $r=2$ meson), the average size of mesons is lower compared to the case when $\theta = 3\pi/4$ where the initial energy is higher. This occurs because the initial kinetic energy also gets converted into the size of mesons.

\begin{figure*}
    \includegraphics[width=18cm]{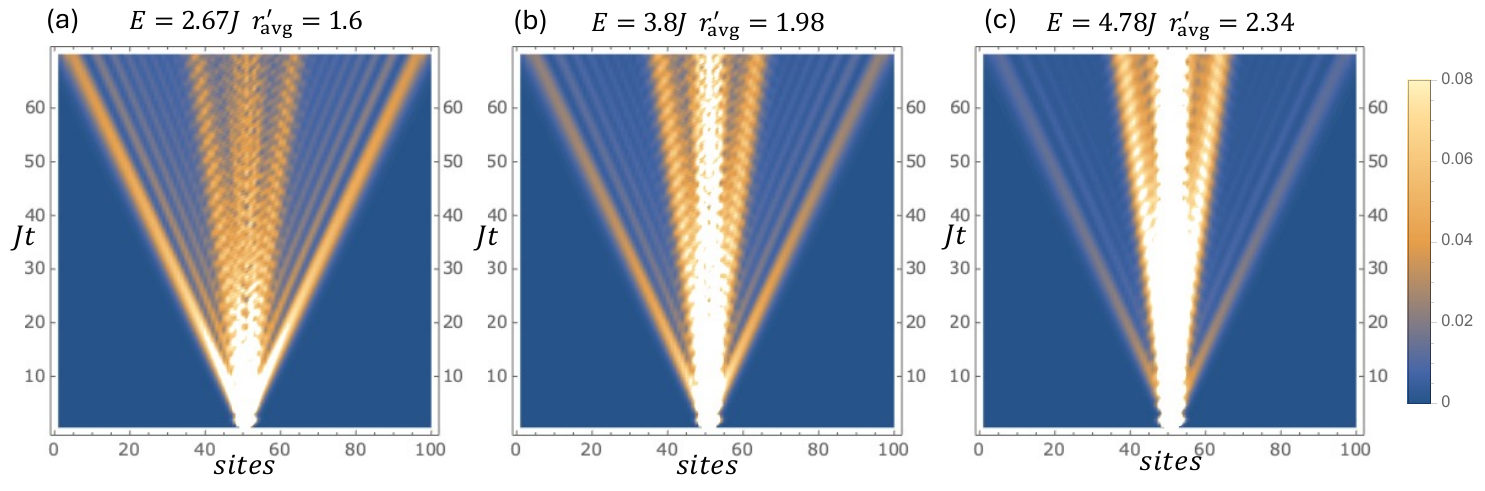}
    \caption{Particle density on each site (x-axis) as a function of time, $Jt$ (y-axis), for $h=1.1J$ for different initial energies, $E$, and average meson size, $r'_{\text{avg}}$. (a) $E=2.67 J$ and $r'_{\text{avg}} = 1.6$, (b) $E=3.8 J$ and $r'_{\text{avg}} = 1.98$ and (c) $E=4.78 J$ and $r'_{\text{avg}} = 2.34$. We can see that as the average meson size, $r'_{\text{avg}}$, increases with energy, the particle density spread becomes narrower. Longer mesons are propagating at slower speeds away from the center.}
    \label{fig:densityenergy}
\end{figure*}

This can be understood from the nature of the eigenstate solutions of the Hamiltonian in Eq.~\ref{hamiltonian2} that we saw in Sec~\ref{sec:twoparticle} in the large $h/J$ limit. The eigenenergy given by $E_{n,k} = 2h n$ is independent of the center-of-mass momentum $k$. For large $h/J$, each eigenfunction corresponding to $E_{n,k}$ is sharply peaked around the relative coordinate value of $r=n$, denoting an $r=n$ meson. Therefore even though the Hamiltonian has contributions from both the  hopping term (kinetic energy) and meson confinement terms (electric field energy), the eigenenergies only depend on the length of the meson excitation.  This property persists even for cases where $h/J < 1$. We have numerically checked that the eigenenergies of the Hamiltonian in Eq.~\ref{wannierstark} are quantized in the bulk of the spectrum (where states with $r \gg 1$ contribute more) for $h/J$ as low as $0.1$. The quantization gets destroyed for small $h/J$ as we near the bottom of the spectrum where $r \sim 1$ states contribute significantly. However, a higher initial energy can place the system closer to the bulk of the spectrum for small $h/J$, somewhat counteracting this lack of quantization. 

\begin{figure}
    \includegraphics[width=\columnwidth]{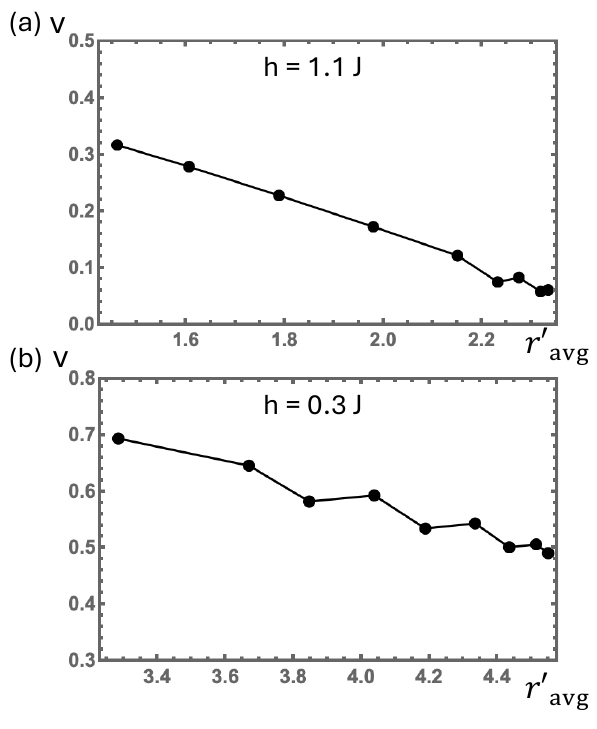}
    \caption{The speed of the propagating meson, $v$, as a function of average meson size, $r'_{\text{avg}}$, at (a) $h=1.1J$ and (b) $h=0.3 J$. The meson size at a fixed $h$ is varied by varying the initial energy of the state. Each data point corresponds to the corresponding data point for angle $\theta$ that varies the initial energy in Fig.~\ref{fig:energysizevstheta}. The speed $v$ and $r'_{\text{avg}}$ are inversely related, making longer mesons slower. This suggests that longer mesons are effectively heavier due to the $Z_2$ gauge constraint. This trend persists for values where $h < J$.}
    \label{fig:speedvssize}
\end{figure}

The speed of the propagating mesons is strongly affected by their size at a fixed confining field, $h/J$. We show the particle density evolution as a function of time for different initial energies and correspondingly different average meson sizes in Fig.~\ref{fig:densityenergy}. We can clearly see that as the average meson size increases, the particle density spread becomes narrower, indicating that longer mesons are propagating at slower speeds. In Fig.~\ref{fig:speedvssize}, we plot the speed of the propagating meson excitations as a function of their size, $r'_{\text{avg}}$, at $h=1.1J$ and $h=0.3 J$ by changing the initial state energy as discussed above. The speed is calculated by finding the slope, $v = dc_s/d(Jt)$ at long times from the $c_s$ vs $Jt$ graph (example shown in Fig.~\ref{fig:dynamics}(b)). We find that meson size and speed are inversely related in both strong confinement ($h > J$) and weak confinement ($h < J$) cases. Longer mesons propagate more slowly. Thus a detector placed far away from the center would detect smaller mesons first and would have to wait for a longer time to detect larger mesons.  This dependence of meson speed on meson size is due to the kinetic constraints arising from the linear confining potential. \textcolor{blue}{This phenomena is known to be nonperturbative in the ratio $h/J$~\cite{bessel}.} Following arguments similar to those in~\cite{fractonconfinement,scattering2}  we show how linear confinement leads to faster speeds for shorter mesons in the $h/J \gg 1$ limit. Furthermore we use a matrix element based approach to quantitatively show how this phenomenology is manifest in the dynamics when $h < J$.

\begin{figure}
    \includegraphics[width=\columnwidth]{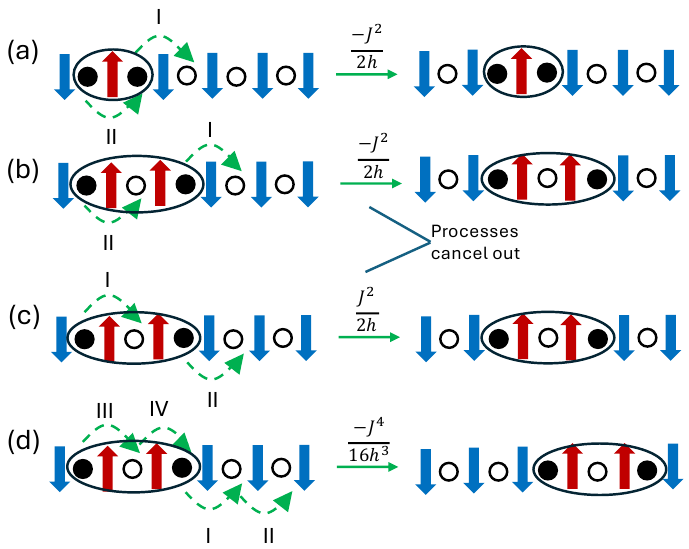}
    \caption{(a) Second-order hopping process (path I followed by path II) of an $r=1$ meson with matrix element $-\textcolor{blue}{J^2}/(2h)$. (b),(c) Two possible second order hopping processes of an $r=2$ meson with path I followed by path II and vice versa. The matrix elements for the two processes are equal and opposite and hence cancel. (d) A fourth order hopping process (I $\rightarrow$ II $\rightarrow$ III $\rightarrow$ IV) with matrix element $-\textcolor{blue}{J^4}/16h^3$ that changes the center of mass position of an $r=2$ meson by 2.}
    \label{fig:hopping}
\end{figure}

We first show how an $r=1$ meson would move faster than an $r=2$ meson and so on for longer mesons when $h > J$. Consider an $r=1$ meson at some large value of $h/J$. The entire meson can hop to its right only by a second order hopping process as shown in Fig.~\ref{fig:hopping}(a). This process changes its center-of-mass position by 1 but leaves the relative coordinate unchanged. By second order perturbation theory, the matrix element for this process is given by $-J^2/2h$. As the confining field $h/J$ increases, the $r=1$ mesons propagate slower. 

Now let's consider an $r=2$ meson. In a similar vein, we can consider processes which lead to the meson hopping to its right. There are two possibilities at second order. The first is shown in Fig.~\ref{fig:hopping}(b) where the right particle hops to the right first and then the left particle follows. The matrix element for this process is $-J^2/2h$ similar to the $r=1$ meson. The second possibility shown in Fig.~\ref{fig:hopping}(c) is similar, but here the left particle hops first, followed by the right particle. The matrix element of this process is $J^2/2h$. The matrix elements of these two processes are equal and opposite in sign. They cancel out and thus the $r=2$ meson cannot hop by a second order hopping process. The leading order process that can cause motion is fourth order, shown in Fig.~\ref{fig:hopping}(d). Here the right particle hops two sites and then the left particle follows. The center of mass changes by 2 and the matrix element for this process is $-J^4/16h^3$. The propagation speed for an $r=2$ meson would be $v \sim J^4/8h^3$ as opposed to $J^2/2h$ for an $r=1$ meson. At large values of $h/J$, the $r=2$ meson would be much slower than the $r=1$ meson. Using a similar argument, one can show that an $r=3$ meson can only propagate by a sixth order process and is even slower than $r=2$ mesons. 

The previous argument seems to hold only when $h/J>1$. However, we will now show that the general argument survives even when $h/J<1$, as seen by slower propagating longer mesons in Fig.~\ref{fig:speedvssize}(b).
Due to the sign cancellations caused by linear confinement, a meson of length $n$ needs at least a $2n^{\text{th}}$ order hopping process to move. The matrix element for such a process, $H_{2n}$, is given by
\begin{align}
    H_{2n} &= -\frac{J^{2n}}{2h \cdot 4h \cdots 2nh \cdots 4h \cdot 2h}\\
 &= J\left(\frac{J}{2h}\right)^{2n-1}\frac{n}{(n!)^2}
\end{align}

\begin{figure}
    \includegraphics[width=\columnwidth]{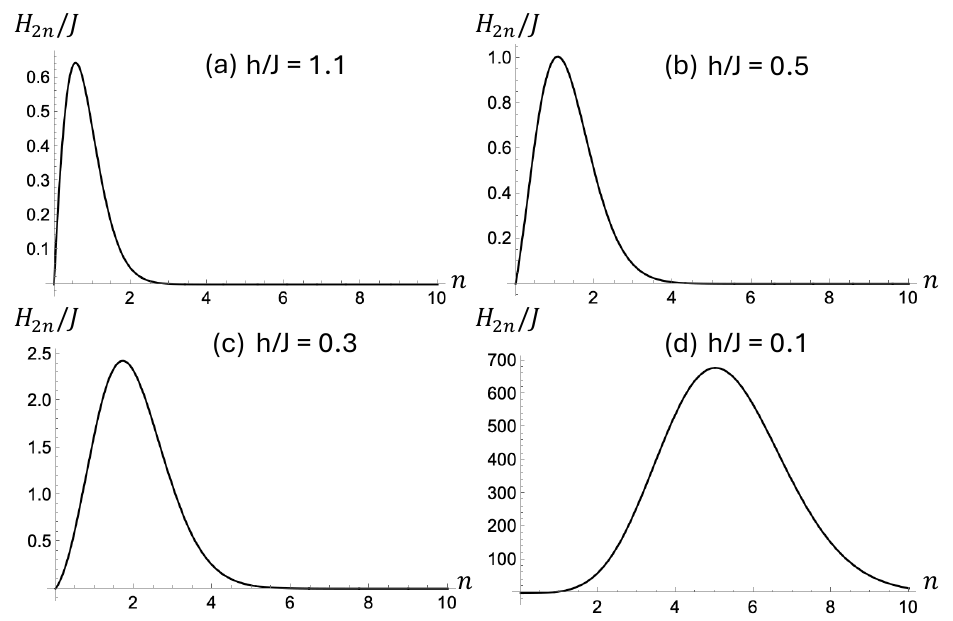}
    \caption{Leading order matrix element, $H_{2n}$ that causes motion of a length $n$ meson by $n$ sites for  (a) $h/J = 1.1$, (b) $h/J = 0.5$, (c) $h/J = 0.3$ and (d) $h/J = 0.1$. For mesons of length beyond the peak $n$, meson length is inversely related to propagation speed since the leading order matrix elements start becoming smaller with size. 
    As $h/J$ decreases, the peak broadens and moves to larger $n$, making the slowdown apparent for only very long mesons.}
    \label{fig:matrixelement}
\end{figure}

We show the matrix element, $H_{2n}$, for $h/J = 0.1, 0.3, 0.5$ and $1.1$ as a function of $n$ in Fig.~\ref{fig:matrixelement}. We can see that for all values of $h/J$ shown here, the matrix element is peaked at some finite $n = n_p$ and larger values of $n>n_p$ ultimately contribute exponentially smaller matrix elements. This is due to the growing $(n!)^2$ factor coming from linear confinement. For mesons shorter than $n_p$, the leading order matrix element is not the largest. Contributions from several possible beyond leading order processes makes the relation of speed and size more complicated. But for mesons longer than $n_p$, an increase in meson size corresponds to a successively lower matrix element even in the leading order process. Thus increase in length would correspond to a decrease in speed.

In Fig.~\ref{fig:speedvssize}, the average meson sizes are greater than the corresponding peak $n_p$ in Fig.~\ref{fig:matrixelement}(a),(c) and thus speed decreases as meson size increases due to increase in energy. We can estimate how the peak $n_p$ varies as a function of $h/J$. Using Stirling's formula for large $n$, we can write,  $H_{2n} \sim e^{2n(\ln |J/h| - \ln n)}$. This function is peaked at $n_p = J/(he)$ where $e$ is the base of the natural logarithm. 
As $h/J$ decreases, $n_p$ increases and the peak broadens. The mesons that propagate slower thus become progressively longer as we approach the continuum limit. The longer mesons slowly propagate out of the central excitation region and damp out the size oscillations at the center as seen prominently in Fig.~\ref{fig:densityplot}(ii).

\section{Proposed experimental implementation using a spin model}

Lattice gauge theories have seen recent implementations in quantum simulator platforms such as cold atoms~\cite{exp1,exp3,exp6}, superconducting qubit-based quantum computer platforms~\cite{exp2,exp5,exp7,exp9,Dalmonte4}, and trapped ions~\cite{stringbreaking3,Gorshkov6}. A promising avenue to implement our particle-conserving $Z_2$ lattice gauge theory model
is by simulating a pure spin model on a quantum computer. The spin model is obtained from the Hamiltonian in Eq.~\ref{hamiltonian} by eliminating the matter fields using the gauge constraint. As shown in ~\cite{sectors,manybody1}, the Hamiltonian maps onto a pure spin model of the form 

\begin{equation}
    H = -J\sum_i \frac{(1-\sigma^z_{i-1,i}\sigma^z_{i+1,i+2})}{2}\sigma^x_{i,i+1} +h\sum_i \sigma^z_{i,i+1}
    \label{spinmodel}
\end{equation}

In the particle vacuum, all the spins point down. A particle excitation corresponds to a domain wall in the spins. The first term in the spin model encodes particle hopping from an occupied site (presence of a spin domain wall) to an empty site (absence of a domain wall) by flipping the link spin in the path. \textcolor{blue}{We can also obtain this model from the infinite Ising coupling limit of the 1D transverse field Ising model. The Ising model Hamiltonian is}

\begin{equation}
    H = -J\sum_i \sigma^x_{i,i+1} +h\sum_i \sigma^z_{i,i+1} + m\sum_i \sigma^z_{i-1,i}\sigma^z_{i,i+1}
    \label{spinmodelising}
\end{equation}

\textcolor{blue}{If the coefficient of the Ising coupling term $m$ is infinitely large, the energy cost to change the number of domain walls is infinite. In dynamics, the number of domain walls (particle excitations) would remain conserved. The effective Hamiltonian in that limit would correspond to the Hamiltonian in ~\ref{spinmodel}. For large but finite $m$, although the number of domain walls are not exactly conserved, we expect our results to continue to hold in the Ising model based on previously shown prethermalization arguments~\cite{bessel,prethermal}.}

\textcolor{blue}{The initial state in the dynamics considered here are simply product states of the spins. Spin product states in the context of lattice gauge theories have been prepared in superconducting circuits~\cite{exp2,exp5,exp7} and trapped ions~\cite{stringbreaking3,Gorshkov6}.} The initial spin configuration for the $r=1$ meson initial state is shown in Fig.~\ref{fig:latticediag}(b). When we tune the initial state energy by introducing a finite amplitude of $r=2$ meson, we simply need to prepare the spin configuration shown in Fig.~\ref{fig:r2meson}. Here the central spin is pointing up and the spin next to it is rotated along the $y$-axis by an angle $\theta$ using a single qubit $y$-rotation gate. 

For studying dynamics, the spin model in Eq.~\ref{spinmodel} can \textcolor{blue}{be} Trotterized on a digital quantum computer. \textcolor{blue}{Alternatively, it can be studied on analog quantum simulators such as trapped ions by implementing the large-coupling limit of the Ising model in Eq.~\ref{spinmodelising}~\cite{stringbreaking3,Gorshkov6,exp7}.}  In the context of digital simulations, the only non-traditional part of the spin model in Eq.~\ref{spinmodel} is the three-spin term encoding particle hopping. \textcolor{blue}{This comes from the gauge-matter coupling term in the lattice gauge theory model and it requires a native three qubit gate for unitary evolution. As this term is composed of Pauli operators, the three qubit term unitary can be decomposed into multiple two-qubit unitaries involving single qubit rotations and two-qubit gates involving at least one of the three qubits~\cite{threequbit1,threequbit2}. We show this in Appendix~\ref{app:fidelity}. Universal quantum computers can implement this term using their universal two-qubit gate set. In Appendix~\ref{app:fidelity}, we discuss the fidelity of two-qubit gates needed as a function of system size, trotter step and evolution time to observe the phenomenology that we study.} 

All the quantities discussed in our work can be accessed by analyzing the snapshots obtained by measuring all the spins at any time step. A particular error-free snapshot would contain two spin domain walls, indicating the two particles. The distance between the domain walls averaged over snapshots gives the average meson size, $r_{\text{avg}}$. The center-of-mass distribution's width, $c_s$ can also be found by recording the width of the distribution of domain wall positions that occur at any time step in the system. \textcolor{blue}{In Appendix~\ref{app:fidelity}, we discuss the number of snapshots needed to observe the meson dynamics phenomena.} 

\section{Summary and Outlook}

We studied the confined mesons that propagate from a localized high energy excitation in a 1D particle-conserving $Z_2$ lattice gauge theory. We considered simple local excitations by adding two matter particles on the gauge field vacuum state and using single spin rotations to tune the energy of the initial excitation. Our scheme provides an alternative to scenarios involving preparation of colliding wavepackets and isolates the effect of confinement field strength and energy of the initial excitation on the nature of propagating excitations. 

We performed exact diagonalization numerics to simulate the dynamics and
found that the long-time state is characterized by the \textcolor{blue}{propagation of a meson that is a superposition of different meson sizes. The meson shows size oscillations in time.}  The average meson size depends inversely on the confinement field while the frequency of oscillations is proportional to the confining field. The meson size oscillations become larger as the confinement field strength decreases. These size oscillations are similar to the particle tracks seen after wavepacket collisions in lattice gauge theories~\cite{Gorshkov4,collider}.
We were able to explain these behaviors analytically by solving an associated Wannier-Stark problem in the large field ($h/J$) limit. 

We also found that as we increase the energy of the initial local excitation, the average meson size monotonically increases, independent of the spatial extent of the initial state. Both the kinetic energy and confinement energy contribution of the initial state gets reflected into the meson size. 

We calculated the speed of the meson propagation as a function of its size and found that a longer mesons is effectively heavier and propagates more slowly.  Longer mesons carry higher gauge field energy excitations giving them a larger effective mass. \textcolor{blue}{Since the long time wavefunction of a single meson contains a superposition of different meson sizes, the correlation between meson size and speed causes a size dependent spatial separation of the superposition constituents.} We found that these results hold not only when $h/J$ is large but also for smaller values of $h/J$. We showed that for any finite linear confining potential, the leading order meson hopping matrix elements show exponential decay for mesons beyond a certain length. This threshold length rises as the confinement field becomes weaker and continues all the way until we approach a continuum limit. Throughout the study, we only consider simple initial states that create a localized high energy excitation region, making these phenomena experimentally accessible in the near-term quantum simulators. \textcolor{blue}{Our initial states offer a clean setting that is less prone to state preparation errors.} 

Although the model studied here is a simple one-dimensional lattice gauge theory, we already see some features that are reminiscent of standard gauge theories. This includes real-time oscillations of particle flavor (meson size oscillations) and effective mass of particles governed by the gauge field excitations they carry. In particular, these features appear even when the confinement field strength is moderate to small and do not require any string breaking or particle pair production processes. The nature of excitations can be tuned by either varying the confinement field strength or the initial state energy, giving us an intuitive way of probing lattice gauge theories on quantum simulators. 

\textcolor{blue}{Our study is amenable to efficient classical simulations due to a restriction to the two particle case in a particle-conserving lattice gauge theory. However we must emphasize that this work's primary goal is to understand some building blocks of high energy phenomena in lattice gauge theory theories in a simple, clean setting that does not require preparation of moving wavepackets. We were able to show that a localized high energy excitation region can give rise to interesting phenomenology of meson propagation and provide a controllable setting for a detailed characterization analysis.}

\textcolor{blue}{Figuring out these building blocks is crucial as we begin to explore similar phenomenology in several closely related models that are hard to simulate classically. This includes models with particle non-conservation, higher dimensional settings and SU($N$) non-abelian lattice gauge theories. Studying the dynamics induced by localized high energy excitations in these scenarios would lead to an interesting interplay where the initial energy can be used to create gauge field excitations or additional matter particles such as mesons or even baryons in SU($N$) gauge theory models. Our work serves as a nascent yet important step in this direction where we can understand the dynamics of high energy phenomena in lattice gauge theories using only simple initial states as opposed to wavepacket collisions.}

\section{Acknowledgments}

VS acknowledges support from the J. Evans Attwell-Welch fellowship by the Rice Smalley-Curl Institute. KRAH acknowledges support from the W. M. Keck Foundation (Grant No. 995764), the National Science Foundation (PHY-1848304), and the Office of Naval Research (N00014-20-1-2695). We thank Sayak Guha Roy and Andrew Long for fruitful discussions.

\appendix

\section{Meson size breathing oscillation function}\label{breathingmode}

Here we analytically calculate the dynamics of the relative coordinate $r$ as a function of time in the large $h/J$ limit. Earlier we showed that the system eigenstates in the large-$h/J$ and high excitation limit are given by $|\psi_{n,k}\rangle = \sum_{c,r}\phi^k_c \gamma^n_r |c,r\rangle$ where $\phi^k_c = e^{ikc}/\sqrt{2L-1}$ and $\gamma^n_r = J_{r-n}(2J \cos(k/2)/h)$. These functions are peaked around $r=n$ with tails decaying as $(2J\cos (k/2)/h)^{|r-n|}$. The eigenenergy $E_{n,k} = 2hn$ is independent of $k$.

For simplicity, we project into a fixed center-of-mass momentum $k$ state, say $k=0$ and then consider the dynamics of the relative coordinate, $r$ starting from some initial value of $r_0$.
We expand the $h=\infty$ eigenfunction $|r_0\rangle$ to first order in $J/h$ to get

\begin{equation}
    |\tilde{r_0}\rangle = |r_0\rangle -\frac{J}{2h}|r_0+1\rangle+\frac{J}{2h}|r_0-1\rangle.
\end{equation}

The eigenenergy of this state is $2hr_0$. We consider an initial state $|\psi(0)\rangle = |r_0\rangle$. We can expand this in terms of the eigenstates $|\tilde{r}\rangle$ to first order in $J/h$ as

\begin{equation}
    |\psi(0)\rangle =|r_0\rangle = |\Tilde{r_0}\rangle -\frac{J}{2h}|\Tilde{r_0}-1\rangle+\frac{J}{2h}|\Tilde{r_0}+1\rangle.
\end{equation}

At time $t$, the state evolves under the Hamiltonian given in the eigenvalue problem in Eq.~\ref{wannierstark}. Up to a global phase factor, it becomes 
\begin{multline}\label{eqbreathing}
    |\psi(t)\rangle =  |\tilde{r_0}\rangle -\frac{Je^{2iht}}{2h}|\tilde{r_0}-1\rangle+\frac{Je^{-2iht}}{2h}|\tilde{r_0}+1\rangle \\
    = (1+\frac{J^2 \cos (2ht)}{2h^2})|r_0\rangle +\frac{J}{2h}(1-e^{2iht})|r_0-1\rangle\\
    - \frac{J}{2h}(1-e^{-2iht})|r_0+1\rangle.
\end{multline}

\subsection{When $r_0 > 1$}\label{appendix1}

We can analytically calculate the spread of relative coordinate, $r$, about $r_0$ as a function of time by calculating $r_s^2 = \langle (r-r_0)^2 \rangle - \langle (r-r_0) \rangle ^2$. Up to leading order, we find that $\langle r \rangle \sim r_0 $ and $\langle (r-r_0)^2 \rangle \sim  2\sin^2(ht)J^2/h^2$. Thus we get the breathing function, $r_s = \sqrt{2}J/h |\sin ht|$. The amplitude of this function sets the meson size spread and captures the $(h/J)^{-1}$ dependence of $r'_{\text{avg}}$ with oscillation frequency, $\omega = 2h$.

\subsection{When $r_0=1$}\label{appendix2}

Due to the boundary condition, we are restricted to $r \geq 1$. Thus we drop any terms of the form $|r_0-1\rangle$ from the first line of Eq.~\ref{eqbreathing}. We can again calculate the average relative coordinate value $\langle r \rangle$ and find that $\langle r \rangle = 1 + J^2/(2h^2) + J^2 \sin^2 (ht)/(2h^2)$. This shows that $r_{\text{avg}}$ oscillates with a frequency, $\omega = 2h$. We also get an $(h/J)^{-2}$ dependence of $r'_{\text{avg}}$.

\section{\textcolor{blue}{Scaling of average meson size with confinement field}}\label{app:scaling}

\begin{figure}
    \includegraphics[width=\columnwidth]{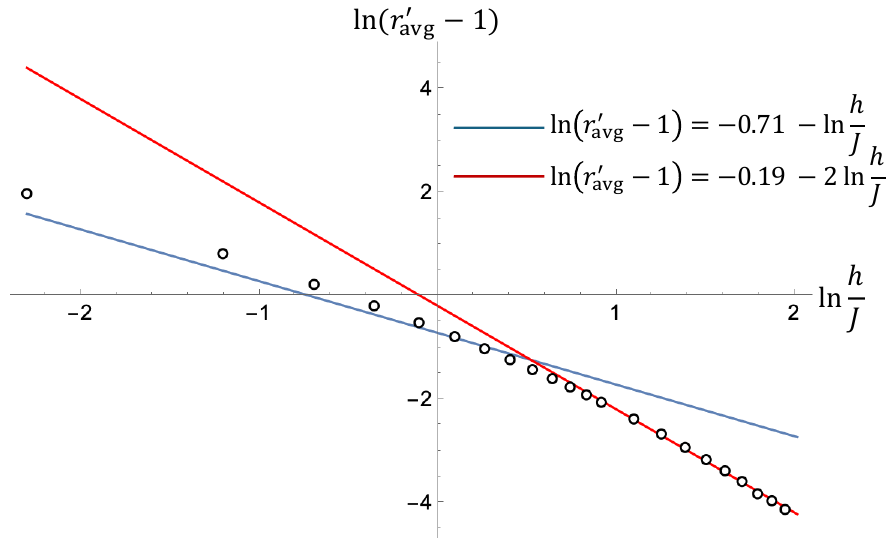}
    \caption{\textcolor{blue}{Long-time average size of the propagating meson, $r'_{\text{avg}}$ vs confinement field, $h/J$ in a log-log plot. The two solid lines are linear fits to the data with the blue and red line having slopes of -1 and -2 respectively.}}
    \label{fig:scaling}
\end{figure}

\textcolor{blue}{Here we show the scaling of the long time average of the meson size, $r'_{\text{avg}}$ as a function of the confinement field, $h/J$. Fig.~\ref{fig:scaling} shows our results. We can see that for moderate values of the confinement fields where $h/J\lesssim 2$, the data follows the blue line showing the scaling, $r'_{\text{avg}} - 1 \propto J/h$. For $h/J > 2$, the scaling,  $r'_{\text{avg}} - 1 \propto (J/h)^2$ shown by the red line is a better fit to the data. This agrees with our analytical prediction in the $h/J \gg 1$ limit that we discussed in Sec.~\ref{sec:fieldeffect}.}

\section{\textcolor{blue}{Two-qubit gate fidelity needed for digital quantum simulation}}\label{app:fidelity}

\textcolor{blue}{Here we discuss how the two-qubit fidelity required to faithfully simulate the dynamics on a digital quantum computer scales as a function of system size, evolution time and trotter step. We focus on the Trotterization of the spin model in Eq.~\ref{spinmodel}. It contains local three-qubit terms of the form $\sigma^z_{i-1} \sigma^x_i \sigma^z_{i+1}$ and single qubit $\sigma^x_i,\sigma^z_i$ terms. Here $i$ denotes site index for the gauge spins. We can use a second order Trotter decomposition where the three-qubit term applied for timestep $J\Delta t$ is sandwiched between single qubit terms applied for timestep $J\Delta t/2$. The error per Trotter step scales as $(J\Delta t)^3$. Thus for a total evolution time of $J T$, the number of trotter steps are $T/\Delta t$ and the total Trotter error (at worst) scales as $J^3 T \Delta t^2$.}

\textcolor{blue}{The number of Trotter steps determine how many two-qubit gates are needed. First we note that only the three-qubit term in the spin model requires application of two-qubit gates. A three-qubit unitary of the form, $\exp({i \theta \sigma^z_1 \sigma^x_2 \sigma^z_3})$ can be written in terms of two-qubit gates and single qubit rotations as}

\begin{multline}
    \exp({i \theta \sigma^z_1 \sigma^x_2 \sigma^z_3}) = e^{i\frac{\pi}{4}\sigma^y_1}e^{i\frac{\pi}{4}\sigma^y_3} \text{CNOT}_{2,3}\text{CNOT}_{1,2}  \\
    \times e^{i\theta \sigma^x_1} \text{CNOT}_{1,2} \text{CNOT}_{2,3}e^{-i\frac{\pi}{4}\sigma^y_3}e^{-i\frac{\pi}{4}\sigma^y_1}.
\end{multline}

\textcolor{blue}{Here $\text{CNOT}_{i,j}$ denotes a CNOT gate between qubits labeled $i$ and $j$. We can see that we require four two-qubit gates to implement one local three-qubit term of the spin model. Given a system size $L$, we would need $4L$ applications of two-qubit terms for each trotter step. For the entire time evolution, we would need $4L (T/\Delta t)$ two-qubit gates. If the fidelity per gate is $p$, the circuit fidelity $\epsilon$ scales (at worst) as, $\epsilon \sim p^{4LT/\Delta t}$.} 

\textcolor{blue}{We would now estimate the fidelity $p$ needed to observe the meson dynamics phenomenology. Our main observables are the average meson size, $r_{\text{avg}}$ and width of the center-of-mass distribution, $c_s$. From Fig.~\ref{fig:dynamics}(a),(c), we can see that $r_{\text{avg}}$ roughly asymptotes as early as $Jt \sim 20$. For $h/J = 1.1,0.3$, Figures~\ref{fig:dynamics}(b),(d) show that $c_s$ is less than 20 at $Jt \sim 20$. Thus a system size of $L \sim 20$ can be used to observe the average meson size for these parameter regimes. If we choose total evolution time $JT = 20$, a Trotter step of $J\Delta t = 0.1$, system size $L = 20$, then the number of two-qubit gates required are 16000. The total trotter error is $\sim 0.2$. For a circuit fidelity of $\epsilon \sim 0.5$, the two-qubit gate fidelity needed is, $p \sim 99.995 \%$. This can be achieved in quantum simulators in the near future. For simulating lower values of $h/J$, we require larger system sizes and longer evolution times. This would need even better two-qubit gate fidelity. We note that post-selection schemes can help in mitigating the errors and allowing access to longer times and larger system sizes.}

\textcolor{blue}{The observables, $r_{\text{avg}}$ and $c_s$ can be obtained from snapshots where all qubits are measured along the $z$-axis. Ideally, each snapshot only contains two spin domain walls. The average meson size, $r_{\text{avg}}(t)$ is simply the distance between domain walls averaged over the snapshots. The distribution of the domain wall locations gives the width of the center-of-mass distribution, $c_s$ as a function of time. The error in computing the averages scales as $1/\sqrt{N}$ where $N$ are the number of snapshots. Taking $N \sim 1000$ snapshots makes the standard error of the mean about $0.03$. This is small enough to capture the average meson size trend seen in Fig.~\ref{fig:sizefreq}. These $ 1000$ shots would be needed for each value of time where one intends to do a measurement for collecting data. For $10$ time points, this amounts to $10000$ measurement shots. }

\section{\textcolor{blue}{Methods}}\label{app:methods}

\textcolor{blue}{Our numerical simulations of the dynamics are carried out in the two-particle sector where the effective Hamiltonian is described by Eq.~\ref{hamiltonian1}. We perform exact diagonalization of this Hamiltonian to exactly simulate the dynamics. A source code file is available in the supplementary files. We use system sizes of $L=100,120$. In the two-particle sector, the Hilbert space is encoded by states of the form, $|i_1,i_2\rangle$ where $i_1,i_2$ are the positions of the two particles and we require $i_1 \neq i_2$. The Hilbert space size grows quadratically with system size as $L \choose 2$.  We calculate the exact time evolved wavefunction after diagonalizing the Hamiltonian and measure observables at each time step in units of $0.1 J^{-1}$.  At any given time, the average meson size, $r_{\text{avg}}$ can be found by calculating the expectation value, $\langle|i_1-i_2|\rangle$ while the width of the center-of-mass distribution is found by calculating the expectation values, $\langle((i_1+i_2)/2)^2\rangle$ and $\langle(i_1+i_2)/2\rangle$.} 

\bibliography{main.bib}

\end{document}